\documentclass[12pt,letterpaper]{article}
\usepackage[utf8]{inputenc}
\usepackage[english]{babel}
\usepackage[vmargin=0.6in,hmargin={1in,1in},headsep=0.3in,headheight=1in]{geometry}
\usepackage{subfiles}
\usepackage{tikz}
\usepackage{fix-cm}
\usepackage{enumitem}
\usepackage{multicol}
\usepackage{amsmath, amsthm, amssymb, amsfonts}
\usepackage{array}
\usepackage{relsize}
\usepackage{xcolor}
\usepackage{lastpage}
\usepackage{blindtext}
\usepackage[linktocpage=true]{hyperref}
\hypersetup{
	colorlinks,
	citecolor=blue,
	filecolor=black,
	linkcolor=blue,
	urlcolor=blue
}
\usepackage{graphicx}
\usepackage{subcaption}
\usepackage{wrapfig}
\usepackage{indentfirst}
\usepackage{hyperref}
\usepackage{soul}
\usepackage{empheq}
\setlist{parsep=0pt,listparindent=\parindent}
\setlist[itemize]{noitemsep, topsep=0pt}
\setlist[enumerate]{noitemsep, topsep=0pt}
\setlist{parsep=0pt,listparindent=\parindent}

\stepcounter{footnote}
\graphicspath{{Images/}{../Images/}}
\catcode`\^=13\def^#1{\sp{#1}{}}
\catcode`\_=13\def_#1{\sb{#1}{}}

\title{\textbf{\Huge{Curvature Invariants for Lorentzian Traversable Wormholes}}}
\author{B.~Mattingly$^{1,2}$\footnote{mailto: \em\texttt{\href{Brandon_Mattingly@Baylor.edu}{Brandon\_Mattingly@Baylor.edu}}}, A.~Kar$^{1,2}$, W.~Julius$^{1,2}$,  M.~Gorban$^{1,2}$, C.~Watson$^{1,2}$, M.~D.~Ali$^{1,2}$, \\ A.~Baas$^{1,2}$, C.~Elmore$^{1,2}$, B. Shakerin$^{1,2}$, E.~W.~Davis$^{1,3}$ and G.~B.~Cleaver$^{1,2}$}
\date{}
\begin{document}
		
\maketitle
\vspace{-0.8cm}
\begin{center}
\begin{minipage}[c]{0.72\textwidth}
	$^1$\emph{Early Universe, Cosmology and Strings \textnormal{(EUCOS)} Group, Center for Astrophysics, Space Physics and Engineering Research \textnormal{(CASPER)}, Baylor University, Waco, TX 76798, USA}\\ \\
	$^2$\emph{Department of Physics, Baylor University, Waco, TX 76798, USA}\\ \\
	$^3$\emph{Institute for Advanced Studies at Austin, 11855 Research Blvd., Austin, TX 78759, USA}\\
\end{minipage}
\end{center}	

\begin{minipage}{0.92\textwidth}
	\textbf{\large{Abstract}:} A process for using curvature invariants is applied as a new means to evaluate the traversability of Lorentzian wormholes and to display the wormhole spacetime manifold. This approach was formulated by Henry, Overduin and Wilcomb for Black Holes in Reference \cite{1}. Curvature invariants are independent of coordinate basis, so the process is free of coordinate mapping distortions and the same regardless of your chosen coordinates. The four independent Carminati and McLenaghan (CM) invariants are calculated and the non-zero curvature invariant functions are plotted. Three example traversable wormhole metrics $\left(i\right)$ spherically symmetric Morris and Thorne, $\left(ii\right)$ thin-shell Schwarzschild wormholes, and $\left(iii\right)$ the exponential metric are investigated and are demonstrated to be traversable.   
	\end{minipage}
	
\section{Introduction}
Lorentzian traversable wormholes were first predicted by Kip Thorne and collaborators who used Einstein's general relativistic field equations to explore the possibility of Faster-Than-Light (FTL) interstellar spaceflight without violating Special Relativity \cite{Morris:1988cz,3}. References \cite{4,5} published earlier studies that demonstrated the possibility of traversable wormholes in general relativity. A Lorentzian traversable wormhole is a topological opening in spacetime which manifests traversable intra-universe and/or inter-universe connections, as well as possible different chronological connections between distant spacetime points. In \cite{6}, M. Visser establishes that a Lorentzian wormhole is traversable provided it is free of both event horizons and singularities. Such a wormhole is fully traversable in both directions, geodesically complete, and there are no crushing gravitational tidal forces found anywhere inside. Consequently, Lorentzian traversable wormholes are unlike the non-traversable Schwarzschild wormhole, or Einstein-Rosen bridge, associated with eternal black holes in the maximally extended version of the Schwarzschild metric. Exotic matter, which violates the point-wise and averaged energy conditions, is required to open and stabilize a Lorentzian traversable wormhole. A comprehensive technical overview of this subject is found in \cite{6}.\\

Studies of Lorentzian traversable wormholes rely on either calculating the elements of the Riemann curvature tensor, $R^i_{jkl}$, to ``observe'' the effects of the wormhole’s spacetime curvature on photons and matter moving through it or by embedding diagrams. However, the $R^i_{jkl}$ cannot be calculated in an invariant manner because they are functions of the chosen coordinates. Thus, analysis of $R^i_{jkl}$ can be misleading  because coordinate mapping distortions may arise as an artifact of the coordinate choice. Embedding diagrams offer a narrow view of the spacetime manifold. In \cite{Morris:1988cz}, the embedding diagrams depicts the wormhole geometry along just an equatorial ($\theta=\frac{\pi}{2}$) slice through space at a specific moment in time. The embedding diagram also only offers a limited view of the physics involved in the wormhole. The best way to illustrate wormhole spacetimes without such issues is to plot their independent curvature invariants to provide proper visualization of any hidden surprises.\\

Christoffel proved that scalars constructed from the metric and its derivatives must be functions of the metric itself and the Riemann tensor and its covariant derivatives \cite{7}. Curvature invariants are scalar products of Riemann, Ricci or Weyl tensors, or their covariant derivatives. Plotting only the curvature invariants, quantities whose value are the same regardless of the choice of coordinates, is the best way to visualize curved spacetime phenomena without distortion.\\

Fourteen curvature invariants have been defined in the literature but the total rises to seventeen when certain non-degenerate cases are taken into account \cite{8}. The set of invariants derived by Carminati and McLenaghan (CM) in \cite{CM} are of lowest degree and contains a minimal independent set for any Petrov or Segre types. In \cite{Santosuosso:1998he}, it is shown that for class B warped product spacetimes only four CM invariants are needed: the first two Ricci invariants, the Ricci Scalar, and the real component of the Weyl Invariant J. Warped products of class B are line elements of the form $ds^2=ds_{\Sigma_1}^2(u,v)+C(x^{\gamma})^2ds_{\Sigma_2}^2(\theta,\phi)$ subject to the restriction $C(x^{\gamma})^2=r(u,v)^2e(\theta,\phi)^2$. Class $B_1$ spacetimes include all spherical, planar, and hyperbolic spacetimes and contain all spacetimes considered in this paper. \\

In \cite{1}, Henry et al. computed and plotted a number of independent curvature invariants for the hidden interiors of Kerr-Newman black holes. They produced visually stunning 3D plots which revealed the surprisingly complex nature of spacetime curvature in Kerr-Newman black hole interiors. Their work motivated the present authors to undertake a similar study for the case of Lorentzian traversable wormholes. Reported here are the computations and 3D plots for three selected Lorentzian traversable wormholes that are described in \cite{6, Boonserm:2018orb}: $\left(i\right)$ the Morris and Thorne (MT) wormhole, $\left(ii\right)$ the thin-shell Schwarzschild wormhole and $\left(iii\right)$ the exponential metric.  The thin-shell flat-face (TS) wormhole was also analyzed, but all of its invariants were trivial as they were identically zero.

\section{Method to Compute the Invariants}
	To find the invariants, the following are required: metric $g_{ij}$, affine connection $\Gamma^i_{jk}$, Riemann tensor $R^i_{jkl}$, Ricci tensor $R_{ij}$, Ricci scalar $R$, trace free Ricci tensor $S_{ij}$ and Weyl tensor $C_{ijkl}$, with the indices $\{i,j,k,l\}$ ranging from $\{0,n-1\}$, where $n$ is the number of spacetime dimensions. Assuming convention, these are defined by:
	\begin{eqnarray}
	\Gamma^i_{jk}&=&\frac{1}{2} g^{il} \left(\partial_j g_{lk}+\partial_k g_{lj}-\partial_l g_{jk}\right),\label{1}\\
	R^i_{jkl}&=&\partial_k\Gamma^i_{jl}-\partial_l\Gamma^i_{jk}+\Gamma^m_{jl}\Gamma^i_{mk}-\Gamma^m_{jk}\Gamma^i_{ml},\label{2}\\
	R_{ij}&=&\partial_k\Gamma^k_{ij}-\partial_j\Gamma^k_{ik}+\Gamma^m_{ij}\Gamma^k_{mk}-\Gamma^m_{ik}\Gamma^k_{mj},\label{3}\\
	R&=&g^{ij}R_{ij}\ ,\label{4}\\
	S_{ij}&=&R_{ij}-\frac{R}{4}g_{ij},\label{5}\\
	C_{ijkl}&=&R_{ijkl}+\frac{1}{2}\left(g_{il}R_{jk}+g_{jk}R_{il}-g_{ik}R_{jl}-g_{jl}R_{ik}\right)+\frac{1}{6}\left(g_{ik}g_{jl}-g_{il}g_{jk}\right)R.\label{6}
	\end{eqnarray}
	The indices may be raised and lowered respectively by applying the inverse metric, $g^{ij}$, or the metric, $g_{ij}$.\\
	
	The formalism to compute the tensors for thin-shell wormholes is outlined in \cite{6}. In brief, two copies of Minkowski flat space on either side of the wormhole's throat are assumed, identical regions from each space are removed, and then separate regions along the boundary are identified. This formalism leads to a well-behaved wormhole, with the throat being located at the identical boundary between the separate regions. In this formalism, the metric is modified to be: \begin{equation}
	g_{ij}(x)=\mathit{\Theta}\left(\eta(x)\right)g^+_{ij}(x)+\mathit{\Theta}\left(-\eta(x)\right)g^-_{ij}(x),
	\label{7} 
	\end{equation} where $g^\pm_{ij}$ is the metric on the respective sides, $\mathit{\Theta}\left(\eta(x)\right)$ is the Heaviside-step function and $\eta(x)$ is the outward pointing normal from the wormhole’s throat. The radius of the wormhole’s throat is located at the point the regions overlap, $x=a$ (that $x\geq a$ is important to note in regards to analyzing divergences). This formalism requires the second fundamental form $K_{\mu\nu}^{\pm}$ for the analysis at the throat to be:
	\begin{equation}
	K_{ij}^{\pm}=
	\pm\begin{pmatrix}
	\ 0&\ \ 0&\ \ 0&\ \ 0\ \ \\[1mm]
	\ 0&\ \ \frac{1}{R_1}&\ \ 0&\ \ 0\ \ \\[1mm]
	\ 0&\ \ 0&\ \ \frac{1}{R_2}&\ \ 0\ \ \\[1mm]
	\ 0&\ \ 0&\ \ 0&\ \ 0\ \ 
	\end{pmatrix},
	\label{8}
	\end{equation}
	where $R_1$ and $R_2$ are the radii of curvature of the wormhole on either side. The thin-shell formalism modifies the Riemann tensor to become:
	\begin{equation}
	R_{ijkl}=-\delta(\eta)\left[k_{ik}n_jn_l+k_{jl}n_in_k-k_{il}n_jn_k-k_{jk}n_in_l\right]+\mathit{\Theta}\left(\eta\right)R_{ijkl}^{+}+\mathit{\Theta}\left(-\eta\right)R_{ijkl}^{-}.\label{9}
	\end{equation}
	where $\delta(\eta)$ is the delta function, $k_{ij}=K_{ij}^+-K_{ij}^-$ is the discontinuity in the second fundamental form, and $n_i$ is the unit normal to the shell. \\
	
	The thirteen different CM invariants are defined in \cite{CM}. The four CM invariants as required by the syzgies in \cite{Santosuosso:1998he} are the Ricci Scalar from \eqref{4}, the first two Ricci invariants, and the real component of the Weyl Invariant J. The remaining invariants are listed below:
	\begin{eqnarray}
	r_1&=& \frac{1}{4} S_a^b S_b^a, \label{10}\\ 
	r_2&=& -\frac{1}{8} S_a^b S_c^a S_b^c, \label{11} \\
	w_2&=& -\frac{1}{8} \bar{C}_{a b c d} \bar{C}^{a b e f} \bar{C}^{c d}{}_{e f}. \label{12}
	\end{eqnarray}
	The full solutions to the wormhole metrics studied herein were found using Wolfram Mathematica\textsuperscript{\textregistered} and are provided in Appendix $B$.

\section{Morris and Thorne Wormhole}
The MT wormhole is defined by a spacetime, which is spherically symmetric and Lorentzian. The spacetime describes the required traversable wormhole geometry. In the standard Schwarzschild coordinates \cite{Morris:1988cz}, the line element is:
\begin{equation}
ds^2=-e^{2\phi^\pm(r)}dt^2+\frac{dr^2}{\left(1-\frac{b^\pm(r)}{r}\right)}+r^2(d\theta^2+sin^2\theta d\varphi^2).\label{25}
\end{equation}
The standard spherical coordinates are used $(r$ : with circumference $=2\pi r;\ 0\leq\theta\leq\pi;\ 0\leq\varphi\leq 2\pi)$, and $ (-\infty< t<\infty)$ is the proper time of a static observer. $\phi^\pm(r)$ is the freely specifiable redshift function that defines the proper time lapse through the wormhole throat. $b^\pm(r)$ is the freely specifiable shape function that defines the wormhole throat’s spatial (hypersurface) geometry. The $\pm$ indicates the side of the wormhole. The throat described by \eqref{25} is spherical. A fixed constant, $r_0$, is chosen to define the radius of the wormhole throat such that $b^\pm(r_0) = r_0$, which is an isolated minimum. Two coordinate patches of the manifold are then joined at $r_0$. Each patch represents either a different part of the same universe or another universe, and the patches range from $r_0\leq r<\infty$. The condition that the wormhole is horizon free requires that $g_{tt}=-e^{2\phi^\pm(r)} \neq 0$ so that $\phi^\pm(r)$ must be finite everywhere \cite{6, Lobo:2017oab}. The use of Schwarzschild coordinates in \eqref{25} leads to more efficient computations of the Riemann and Ricci curvature tensors, the Ricci scalar, and all four invariants. \\

Using Wolfram Mathematica\textsuperscript{\textregistered}, all four independent invariants were computed and are recorded in \eqref{eq:B.1} through \eqref{eq:B.4}. All the invariants are non-zero and depend only on the radial coordinate, $r$, implying they are spherically symmetric. The invariants are plotted in Figure $1$ after selecting $\Phi(r)=0$ for the redshift function and selecting the shape function
\begin{equation}
b(r)=2GM\left(1-e^{r_0-r}\right)+r_0e^{r_0-r}. \label{26}
\end{equation} 
These functions satisfy the constraints required by \cite{6} as discussed in Appendix $A.2$. At a distant greater than $0.5 r_0$, all the figures are asymptotically flat, which corresponds to zero curvature. For $r\to0$, the figures diverge to infinity. The divergence at $r=0$ is not pathological as the radial coordinate $r$ has a minimum $r_0>0$ at the wormhole’s throat. Thus, a traveler passing through the wormhole would not experience any divergence. Any tidal forces on the traveler would be minimal. Consequently, the MT wormhole would be traversable as indicated by the included invariant plots. \\

\begin{figure}[h]
	\centering
	\begin{subfigure}[t]{.45\linewidth}
		\includegraphics[scale=0.3]{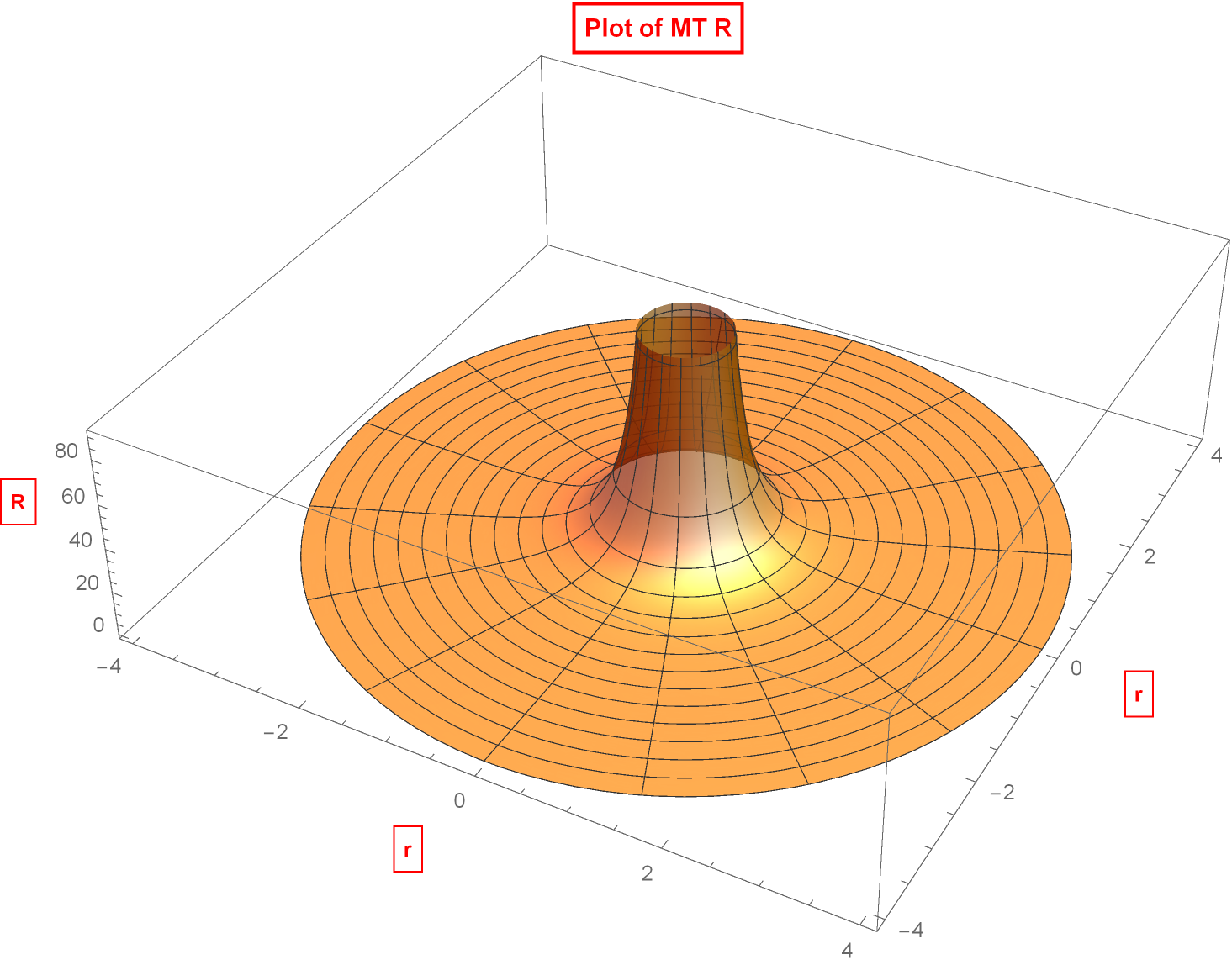}
		\caption{Plot of MT $R$}
		\label{RMTm}
	\end{subfigure}
	~
	\begin{subfigure}[t]{.45\linewidth}
		\includegraphics[scale=0.3]{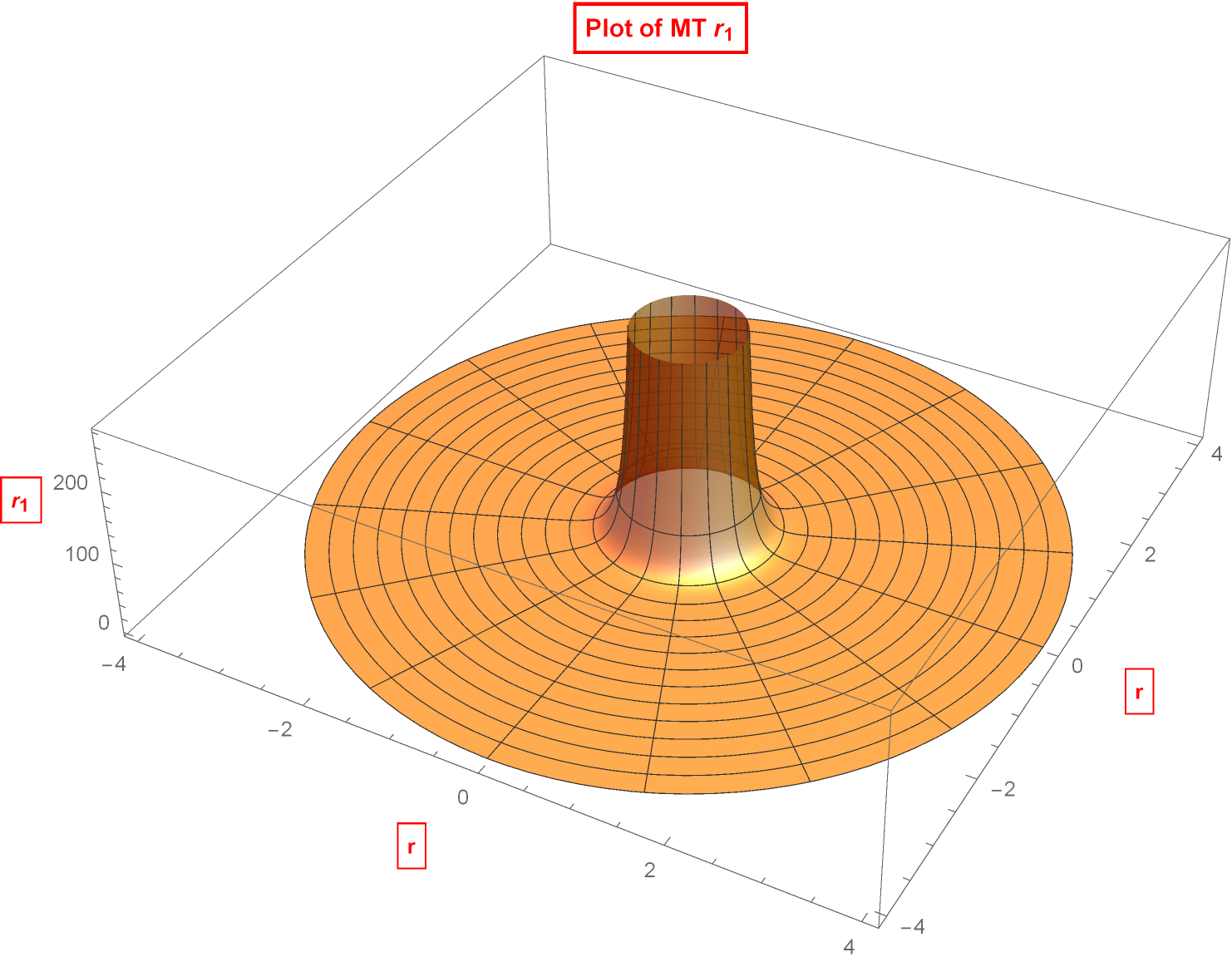}
		\caption{Plot of MT $r_1$}
		\label{r1MTm}
	\end{subfigure}
	~
	\begin{subfigure}[t]{.45\linewidth}
		\includegraphics[scale=0.3]{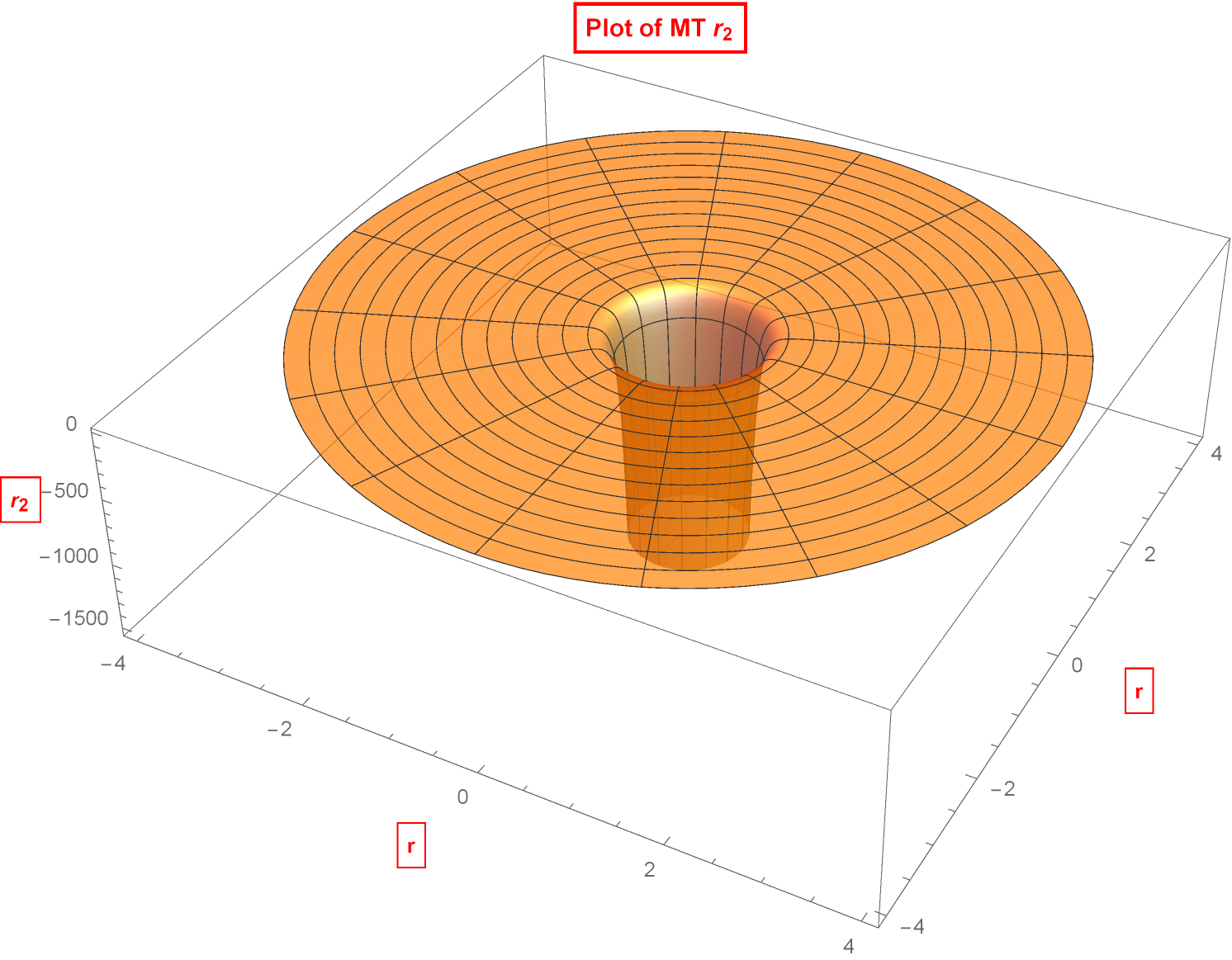}
		\caption{Plot of MT $r_2$}
		\label{r2MTm}
	\end{subfigure}
	~
	\begin{subfigure}[t]{.45\linewidth}
		\includegraphics[scale=0.3]{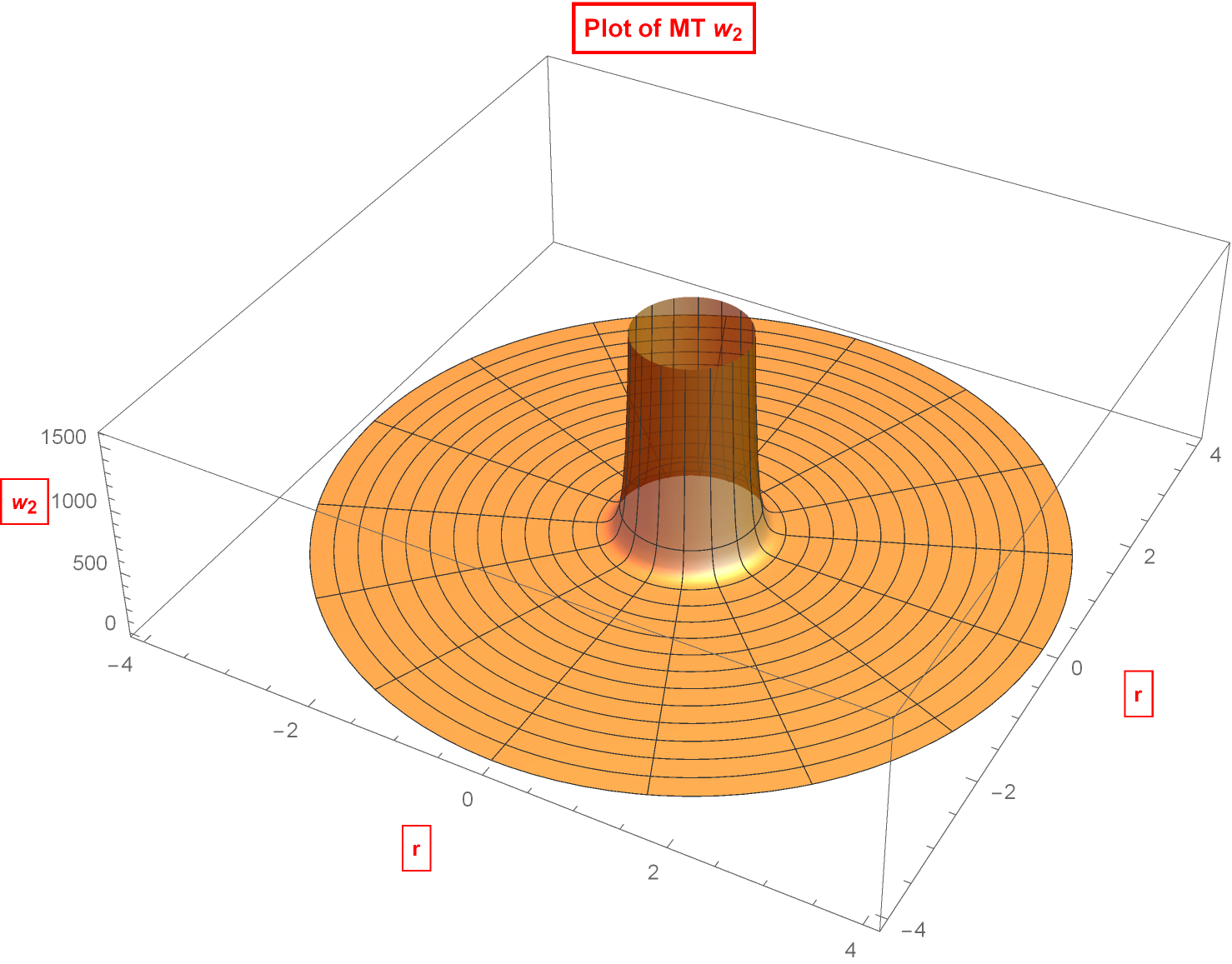}
		\caption{Plot of MT $w_2$}
		\label{w2MTm}
	\end{subfigure}
	\caption{Plots of the non-zero invariants for the MT wormhole. The plots are in radial coordinates with $r\in\{0,4\}$. $G=M=1$ were normalized for simplicity and $r_0=2$ was chosen as the throat. Notice the divergence at the center of each plot is completely inside the $r=2=r_0$ radial line. This does not affect the traversability of the wormhole. }
\end{figure}

\newpage

\section{Thin-Shell Schwarzschild Wormhole}
A second example wormhole is the Thin-Shell Schwarzschild wormhole. It is constructed by the steps described in \cite{6}. The Schwarzschild geometry in natural units is given by the line element: 
\begin{equation}
ds^2=-\left(1-\frac{2M}{r}\right)dt^2+\frac{dr^2}{\left(1-\frac{2M}{r}\right)}+r^2\left(d\theta^2+sin^2\theta d\varphi^2\right),\label{27}
\end{equation}
where $M$ is the mass of the wormhole. The standard spherical coordinates are used. The thin-shell formalism is applied with a unit normal $n_i=\left(0, \sqrt{1-\frac{2M}{r}}, 0, 0\right)$ in the notation of \cite{6}. Regions described by $\Omega_{1,2}\equiv\left\{r_{1,2}\leq a\ |\ a>\frac{3M}{2} \right\}$ are removed from the two spacetimes leaving two separate and incomplete regions with boundaries given by the time-like hypersurfaces $\partial\Omega_{1,2}\equiv\left\{r_{1,2}=a\ |\ a>\frac{3M}{2} \right\}$. The boundaries $\partial\Omega_{1}=\partial\Omega_{2}$ at the wormhole throat of $r=a$ are identified and connected.  The boundary at $a=\frac{3M}{2}$ is chosen to satisfy the Einstein equations and equation of state in \cite{6}; however, an event horizon is expected. The resulting spacetime manifold is geodesically complete and contains two asymptotically flat regions connected by the wormhole.\\

The four curvature invariants were computed for the Schwarzschild wormhole using Wolfram Mathematica\textsuperscript{\textregistered} and are recorded in \eqref{eq:B.5} and \eqref{eq:B.6}. The mass and radius of the throat are normalized to $M=1$ and $a=\frac{3}{2}$. All curvature invariants vanish with the exception of $w_2$. The $w_2$ invariant is broken into two main portions. The first part is $\frac{-12M^3}{r^9}$, which equals the $w_2$ invariant from the Schwarzschild black hole line element. The remaining portions of the function are all proportional to different powers of $\delta\left(r-a\right)$. This portion is of interest as it is on the throat between the hypersurfaces. \\

The only nonzero invariant, $w_2$ is plotted in Figure $2$. Its plot has one divergence and one discontinuity. The divergence occurs at $r=0$, which is outside the manifold of $\Omega_{1,2}$. By the same argument for the apparent MT divergence, the first Schwarzschild divergence would not impede the traversability of the wormhole. The discontinuity occurs at $r=a=\frac{3M}{2}$ and is located at the throat where the horizons are connected by the Schwarzschild wormholes. In these invariants, it is represented by a discontinuous jump to the value in \eqref{eq:B.7}. Since the invariants at the horizon are inversely proportional to $a^{-14}$, the tidal forces on a traveler is benign at the horizon, and the Thin-Shell Schwarzschild wormhole would be traversable. \\
\vspace{3mm}
\begin{figure}[h]
	\centering
	\includegraphics[scale=0.3]{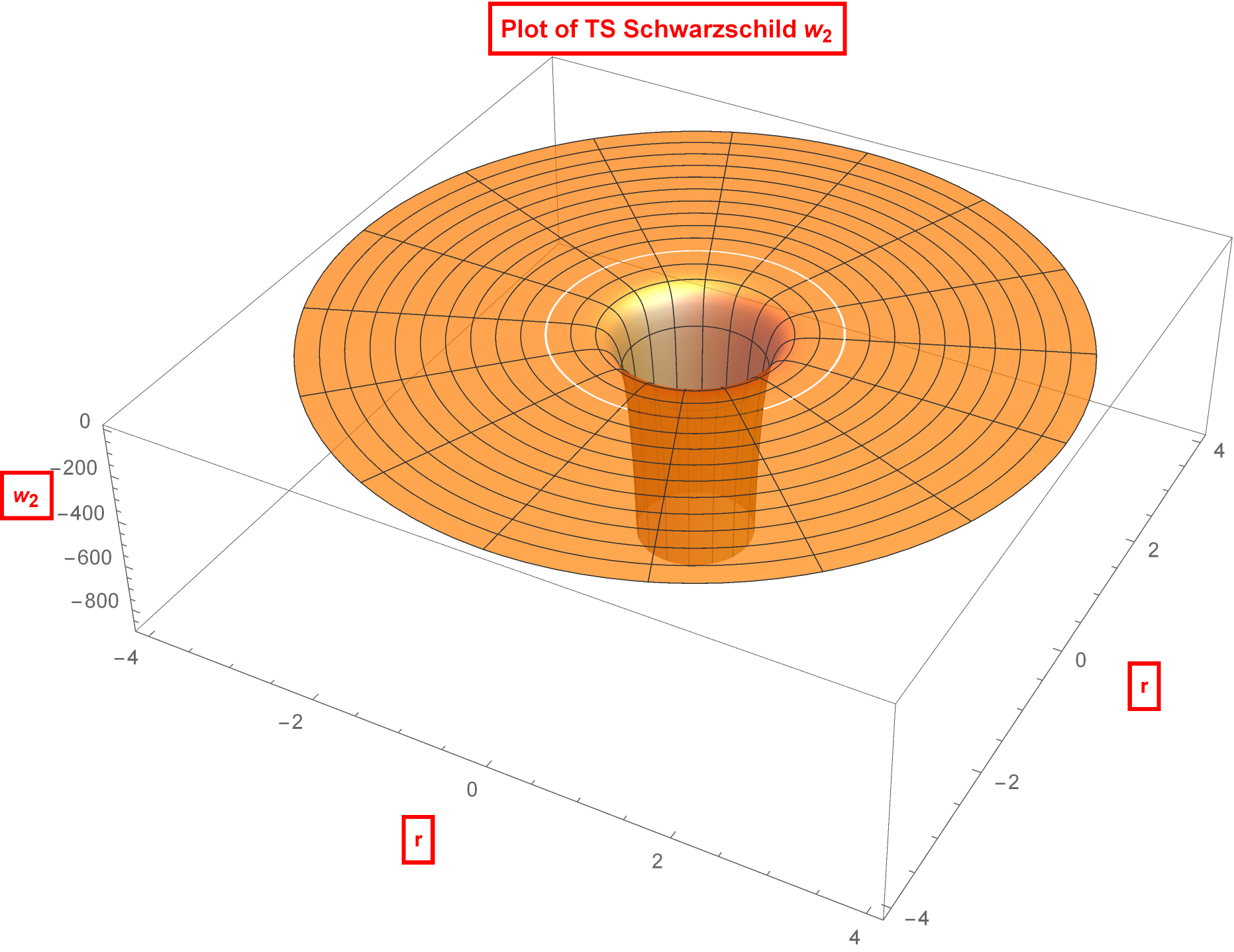}
	\caption{Plot of Schwarzschild $w_2$. The plot is in radial coordinates with $r\in\{0,4\}$. The divergence at the center of the plot is completely inside the $r=2=r_0$ radial line. It will not affect the traversability of the wormhole. In addition, notice the ring discontinuity at $r=2=r_0$. It is from the $\delta$-function in the thin-shell formalism and its value is recorded in \eqref{eq:B.7}.}
	\label{w2tss}
\end{figure}

\newpage
 
\section{The Exponential Metric}
The exponential metric was demonstrated recently in \cite{Boonserm:2018orb} to have a traversable wormhole throat. In natural units, its line element is
\begin{equation}
ds^2=-e^{\frac{-2M}{r}}dt^2+e^{\frac{+2M}{r}}\{dr^2+r^2\left(d\theta^2+sin^2\theta d\varphi^2\right)\},\label{30}
\end{equation}
where M is the mass of the wormhole and the standard spherical coordinates are used. 
It has a traversable wormhole throat at $r=M$. The area of the wormhole is a concave function with a minima at the throat where it satisfies the ``flare out'' condition. It does not have a horizon since $g_{tt}\neq0$ for all $r\geq0$. The region $r<M$ on the other side of the wormhole is an infinite volume ``other universe'' that exhibits an ``underhill effect'' where time runs slower since $e^{\frac{-2M}{r}}>0$ in this region. \\

The four curvature invariants were computed for the exponential metric using Wolfram Mathematica\textsuperscript{\textregistered}. They are recorded in \eqref{eq:B.8} through \eqref{eq:B.11} and plotted in Figure $3$. They are all nonzero and depend only the radial coordinate $r$ implying spherical symmetry. In addition, they are finite at the throat $r=M$ and go to zero as $r\xrightarrow{}\infty$ in accordance with \cite{Boonserm:2018orb}. $w_2$ and $R$ have a minima near the throat, while $r_1$ and $r_2$ have a maxima. The plots are finite everywhere and completely connected confirming the lack of a horizon. The encountered tidal forces would be minimal. It can be concluded that the exponential metric represents a traversable wormhole.

\begin{figure}[h]
	\centering
	\begin{subfigure}[t]{.45\linewidth}
		\includegraphics[scale=0.50]{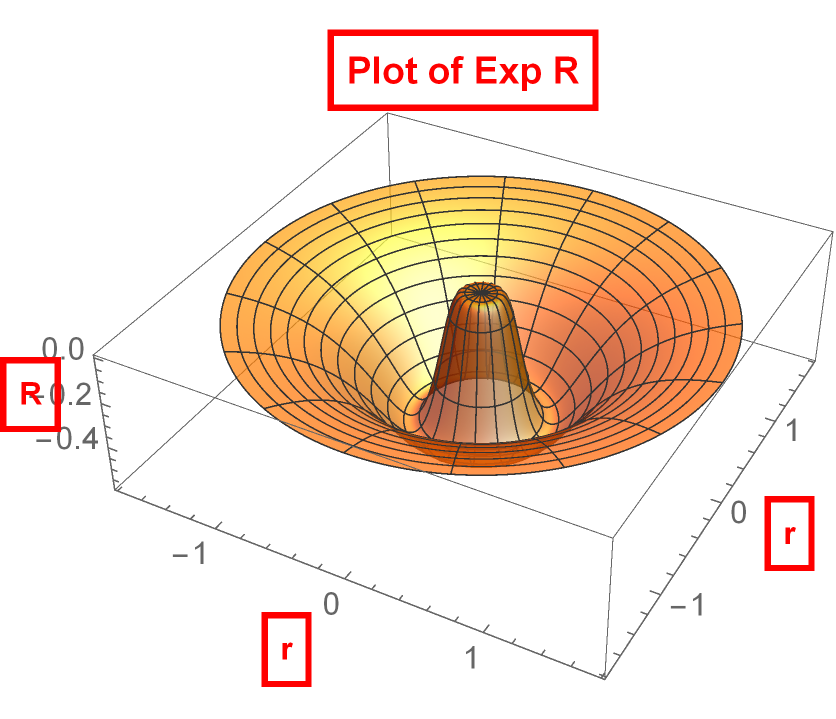}
		\caption{Plot of the exponential metric $R$}
		\label{Rem}
	\end{subfigure}
	~
	\begin{subfigure}[t]{.45\linewidth}
		\includegraphics[scale=0.50]{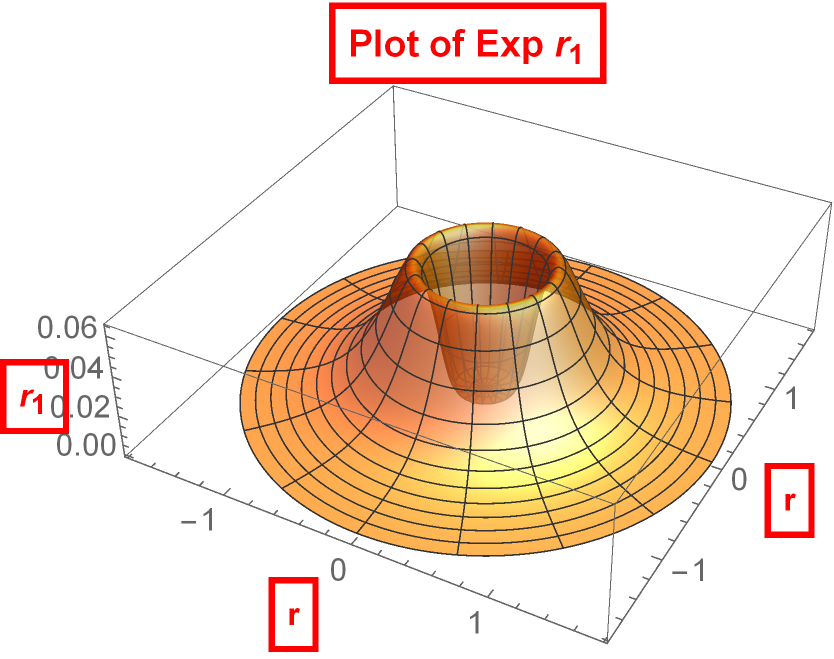}
		\caption{Plot of exponential metric $r_1$}
		\label{r1em}
	\end{subfigure}
	~
	\begin{subfigure}[t]{.45\linewidth}
		\includegraphics[scale=0.50]{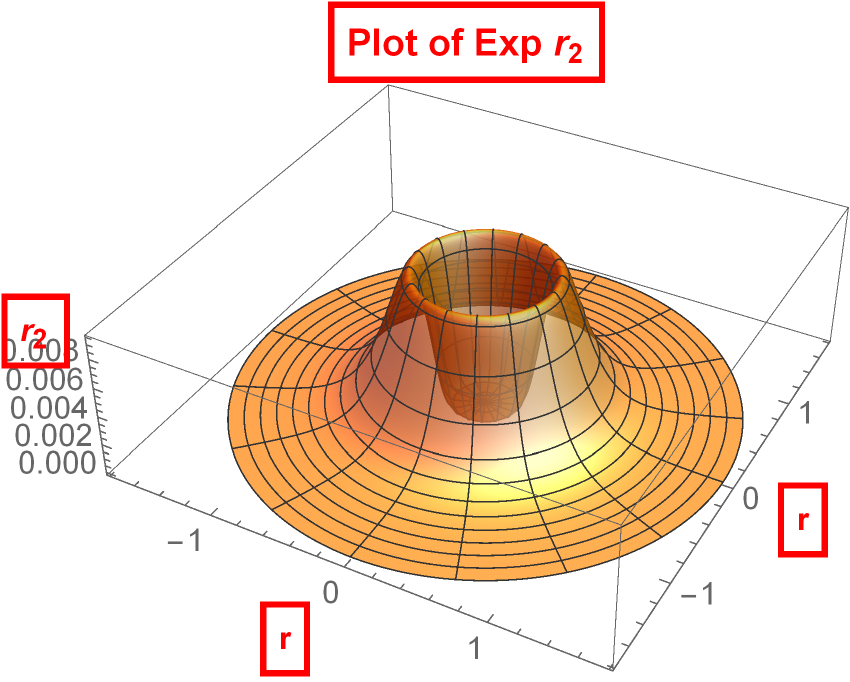}
		\caption{Plot of exponential metric $r_2$}
		\label{r2em}
	\end{subfigure}
	~
	\begin{subfigure}[t]{.45\linewidth}
		\includegraphics[scale=0.50]{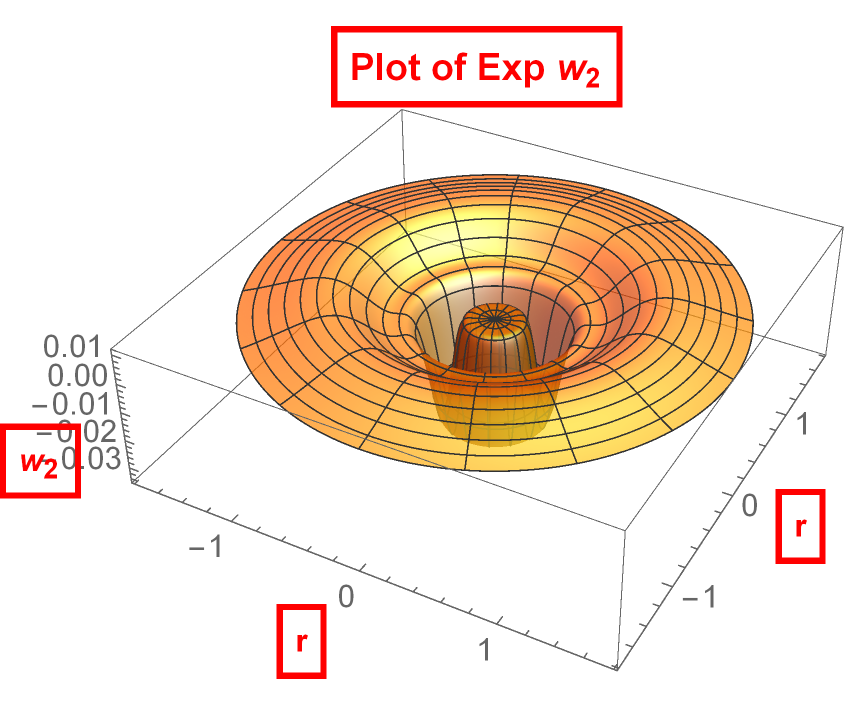}
		\caption{Plot of the exponential metric $w_2$}
		\label{w2em}
	\end{subfigure}
	\caption{Plots of the non-zero invariants for the exponential metric. The plots are in radial coordinates with $r\in\{0,1.75M\}$. $M=1$ was normalized. The throat is then at r=1.}
\end{figure}

\newpage

\section{Conclusion}
This paper demonstrates how computing and plotting the curvature invariants of various wormholes can reveal the entire wormhole spacetime manifold and whether the wormhole is traversable or not.  As examples, it is indicated that $\left(i\right)$ spherically symmetric MT, $\left(ii\right)$ thin-shell Schwarzschild and $\left(iii\right)$ exponetial metric wormholes are traversable in agreement with \cite{Morris:1988cz,3,6}. The scalar polynomial invariants of the MT wormhole were found to be non-zero and are plotted in Figures \ref{RMTm}-\ref{w2MTm}. A divergence is found in all four, but it does not affect the wormhole's traversability since the divergence is outside the physical range of the radial coordinate, $r\in{\left(r_0,\infty\right)}$. For the thin-shell Schwarzschild wormhole, $w_2$ is found to be the single non-zero invariant. As plotted in Figures \ref{w2tss}, it has a divergence at the center and a ring discontinuity. The divergence is outside the physical radial coordinate and can be safely ignored. The ring discontinuity represents a jump due to the $\delta$-function from the thin-shell formalism. It is shown to be inversely proportional to ${a^{-14}}$, not affecting traversability through the wormhole. The scalar polynomial invariants of the exponential metric were found to be non-zero and were plotted in \ref{Rem}-\ref{w2em}. The plots are continuous across the entire manifold and traversable.\\

Potentially, the ring discontinuity in the thin-shell Schwarzschild wormhole could lead to a redshift of light rays passing through the wormhole. The redshift could be used to distinguish wormholes from black holes. While significant research remains to answer traversable wormhole issues, especially with regards to the exotic mass requirements and understanding averaged null energy condition violations, the present paper hopes to establish several methods for understanding the traversal through wormholes. \\

Computing and plotting the invariant functions has significant advantages for the inspection of wormholes. As mentioned previously, the advantage of plotting the invariants is that they are free from coordinate mapping distortions and other artifacts of the chosen coordinates.  The resulting invariants properly illustrate the entire underlying spacetime independent of the coordinate system chosen. Plotting the invariants exposes the presence of any artifacts, divergences or discontinuities anywhere on the manifold. Once the artifacts are revealed by the invariants, they can be related mathematically to the tensors in \eqref{1}-\eqref{6}. Their effect on an object's motion can then be analyzed. A second advantage is the relative ease with which the invariants can be plotted. Software packages exist or can be developed to calculate the tensors in \eqref{1}-\eqref{6}. The aforementioned tensors lead to a chosen basis of invariants. While the CM invariants were chosen to be computed and plotted in this paper, other choices exist, such as the Cartan invariants and the Witten and Petrov invariants \cite{8, McNutt:2017paq}. These can be computed and plotted without difficulty. Since the invariants are either scalars or pseudoscalars, they can be straightforwardly plotted and visually interpreted.   \\

The form of the invariant functions and plots for other wormhole metrics can be hypothesized based on the three wormhole metrics studied in this paper. The thin-shell flat-face wormhole chosen is the simplest of a large class of thin-shell wormholes which include spherical, cubic, and polyhedral types. Second fundamental forms for thin-shell wormholes depend on the two radii of curvature, $R_1$ and $R_2$, of the throats. As a consequence of the $\delta$-functions in the thin-shell formalism, these radii would likely lead to jump discontinuities in the general thin-shell invariants similar to those found in the thin-shell Schwarzschild wormhole. Another direction for thin-shell wormholes study is to investigate a more realistic metric on either side of the wormhole such as the Friedmann metric. Another variation of thin-shell wormholes is to model different radii of curvature on either side of the wormhole. Thin-shell wormholes with unequal radii can be further extrapolated to wormholes with different chronologies or universes on either side of the throat.\\

The thin-shell Schwarzschild wormhole is also the most common example of a large class of wormholes. The class includes wormholes with different radii of curvature and/or masses on either side of the throats, wormholes with same or different charge, Q, on either side of the wormhole, and time-dependent wormholes. For charged wormholes, a second ring artifact at $r=Q$ is likely to exist since the metric has a singularity at that point.\\

A prospective future application of this work is an investigation of dynamic wormholes. Dynamic wormholes are the ones where the radii of the two throats change over time. This implies that the ring discontinuity in the invariant functions will change as a function of time. Hence, dynamic wormholes are technically more intricate to study as compared to static wormholes.  Consequently, it can be expected that the computation of a dynamic wormhole's invariants and their plots increase in difficulty and computational runtime. In a broader perspective, the calculation and plotting of curvature invariants can be made to encompass other types of FTL spaceflight such as the Alcubierre warp drive \cite{12}.
\newpage

\section{Acknowledgements}
E.~W.~Davis would like to thank the Institute for Advanced Studies at Austin for supporting this work. B.~Mattingly would like to thank D.~D.~McNutt for beneficial discussions. 		

\appendix
\section{Appendices}
\subsection{Riemann Curvature Invariants}
Riemann curvature invariants are scalar products of the Riemann, Ricci, and Weyl tensors and their traces, covariant derivatives, and/or duals. Invariants are measures of curvature, which is defined as the amount by which the spacetime geometry differs from being flat \cite{8}. The prime example of the invariants are scalar polynomial (SP) invariants such as the Kretschmann invariant, $R^{ijkl}R_{ijkl}$, though other types exist, such as the Cartan invariants. This paper focuses on the SP invariants (the SP prefix should be assumed). In the invariants, Einstein summation is performed over repeated indices, resulting in a scalar function formed from various polynomials. Reference \cite{8} notes that the complete set of invariants are important in studying general relativity since they allow a manifestly coordinate invariant characterization of certain geometrical properties of spacetimes. Invariants are critical for studying curvature singularities, the Petrov type of the Weyl tensor and the Segre type of the trace free Ricci tensor, and for studying the equivalence problem\footnote{The equivalence problem is whether two different metrics lead to identical spacetimes \cite{8}.}. In this paper, the invariants are primarily used to study the curvature singularities.  \\

A simple example of a sphere given in \cite{9} can help illustrate the use of invariants. The metric of a 2-sphere is 
\begin{equation}
g_{ij}=
\begin{pmatrix}
\ a^2&\ \ 0&\ \\
\ 0&\ \ a^2 \sin^2{\theta}& 
\end{pmatrix}.\tag{A.1} \label{A.1}
\end{equation}
where a is the radius of the sphere. There are two nonzero components of the Riemann tensor, $R^1_{221}=\sin^2{\theta}$ and $R^1_{212}=\sin^2{\theta}$, computed from \eqref{2}, which fully determine the curvature of the sphere. However, normally we think of the curvature in terms of the Gaussian curvature computed from \eqref{4}. For the sphere, $R=\frac{1}{a^2}$, which is related with the circle bounding the equator. Alternatively, any other invariant for other characteristics of the curvature can be computed. For example, the Kretschmann invariant gives $R^{ijkl}R_{ijkl}=\frac{2}{a^4}$, which can be connected with the surface area of the sphere. Here, the curvature invariants measure the curvature of the manifold and not the object's path through the manifold. Unfortunately, the invariants in Appendix B are not as simple as the invariants of the sphere. To gain a physical insight into the nature of the invariants, they must be plotted as is done herein. \\

\newpage 
The CM invariants are divided into three groups: the Weyl, the Ricci, and the mixed. In general, there are four independent Weyl invariants given by the real and complex parts of functions defined as $I$ and $J$ in \cite{8}. IN this paper, $w_2$ is a member of the Weyl invariants. As for the Ricci invariants, there are four independent real Ricci invariants formed from the Ricci scalar and three traces of the Riemann, Ricci, or trace-free Ricci tensors. In this paper, $R$, $r_1$, and $r_2$ are members of the Ricci invariants. There are at most six different mixed invariants formed by combining combinations of the Riemann, Ricci, or trace-free Ricci tensors.  \\

As discussed in \cite{11}, singularities come in three types in general relativity: coordinate, removable, and intrinsic. Coordinate singularities result from a coordinate system only covering a portion of the manifold. The classic example is Schwarzschild coordinates used in \eqref{25} and \eqref{26}, which do not cover the axis at $\theta=\{0,\pi\}$ because the line element becomes degenerate and the metric ceases to be of rank 4. Coordinate singularities are removable by a proper change of coordinates and will not appear in the invariants. As illustrated in Figure \ref{w2tss}, the coordinate singularity does not manifest itself in the plots as expected.  \\

The removable type of singularity can be apparently seen in the metric, but vanishes in any calculated invariants. As an example, the singularity in the metric in \eqref{27} at $r=2M$ is removable. Upon inspection of the invariants in \eqref{eq:B.10} and \eqref{eq:B.11}, the $r=2M$ singularity is removed by replacement with a $\delta$-function. In Figure \ref{w2tss}, the singularity is absent.  \\

The final type of singularity is an intrinsic (a.k.a.~a curvature, physical, essential, or real) singularity. This type of singularity can not be removed by any proper change of coordinates and remains a singularity in the invariants. An example of an intrinsic singularity is the $r=0$ singularity seen in \eqref{eq:B.3} - \eqref{eq:B.8}, \eqref{eq:B.10}, and \eqref{eq:B.11}. This can be seen in each of the plots: \ref{RMTm} - \ref{w2MTm}, and \ref{w2tss}. However, it should be noted that an invariant blowing up at a point does not necessarily imply a singularity exists in the spacetime manifold. For example, this paper defined its coordinates in a way not to pass through the $r=0$ singularity in any of the spacetimes considered herein. 
\newpage
\subsection{The Morris and Thorne Redshift and Shape Functions}
The  MT Wormhole given in \eqref{25} has two freely specifiable functions i.e., the redshift function, $\phi_\pm(r)$, and the shape function, $b_\pm(r)$. These two functions satisfy the consistency requirements as \cite{6} entails. \\

The redshift function must have:
\begin{enumerate}
	\item continuity of the $t$ coordinate across the throat, $\phi_+(r_0)=\phi_-(r_0)$,
	\item existence and finiteness of both the limits, $\lim_{r\to\pm\infty} \phi(r)=\phi_\pm$.
\end{enumerate} 
These two aforementioned constraints are the minimum required for a traversable wormhole. However, additional constraints can be imposed by choice on the redshift function for ease of calculations and temporal and spatial symmetry.  \\

The shape function must have:
\begin{enumerate}
	\item existence and finiteness of both the limits, $\lim_{r\to\pm\infty} b(r)=b_\pm$,
	\item the masses of the wormhole $M_\pm$ on the two sides are given by $b_\pm=2GM_\pm$,
	\item $\exists r_*\|\forall r\in(r_0,r*),b'(r)<\frac{b(r)}{r}$,
	\item $b_+(r_0)=b_-(r_0)$ and $b'_+(r_0)=b'_-(r_0)$ at the throat. 
\end{enumerate} 
The shape function chosen in \eqref{26}, $b(r)=2GM\left(1-e^{r_0-r}\right)+r_0e^{r_0-r}$, obeys the listed conditions, with $\lim_{r\to\pm\infty} b(r)=2GM$. Thus, the shape function at the throat exists, is finite, and is continuous, which satisfies the second and the fourth conditions. At the throat (i.e., $r\to r_0$), $b(r)=r_0$. The derivative, $b'(r)=(2GM-r_0)e^{r_0-r}=b(r)+2GM$, is in agreement with the third condition mentioned above\footnote{Since $G=1$ in natural units and $M>0$, $b'(r)<\frac{b(r)}{r}$ for all $r_0\le r\le \infty$ instead of a specific range of $r_*$ as necessitated by the third condition.}. \\

\begin{figure}[h]
	\centering
	\includegraphics[scale=0.3]{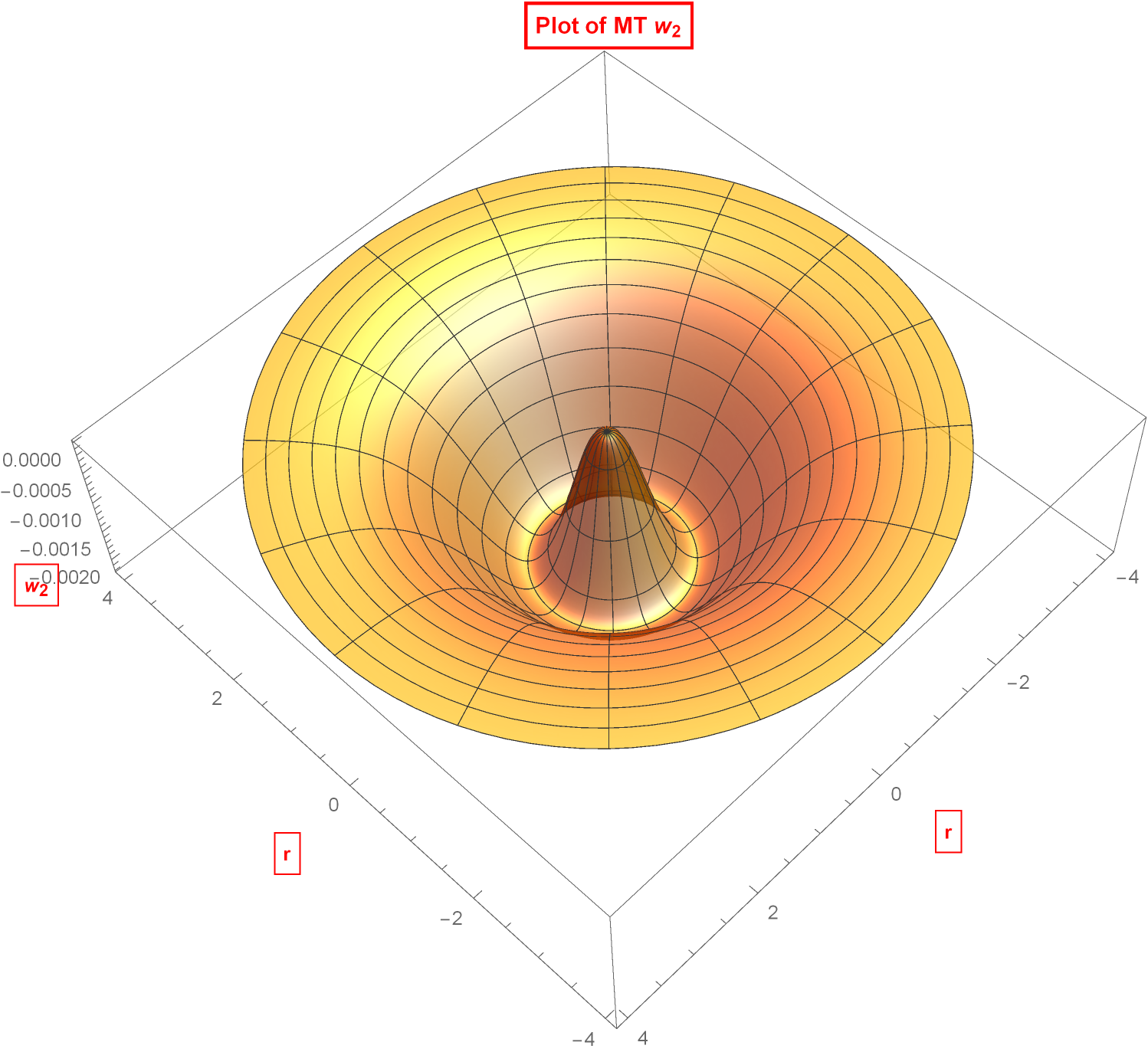}
	\caption{Plot of MT $w_2$ for the shape function given in \eqref{eq:A.2}. The plots are in radial coordinates with $r\in\{0,4\}$ with $G=M=1$ normalized and $r_0=2$ chosen.}
	\label{ewMTm}
\end{figure}

The freely specified shape function can have a significant impact on the form of the invariant functions. By inspecting the invariants in Appendix B, it is seen that a shape function with a term of $r^n$ with $n\ge3$ will not have a discontinuity at $r=0$. For example, a shape function 
\begin{equation}
b(r)=\frac{r^3}{r_0^3}\left(e^{r_0-r}\right), \tag{A.2} \label{eq:A.2}
\end{equation}
gives a plot of $w_2$ in Figure \ref{ewMTm}. The discontinuity at $r=0$ has been replaced by a trench and an inverted cone with a finite depth that resembles the Ricci Scalar from the exponential metric shown in Figure \ref{Rem}. While this is outside the domain of $r$ as discussed in $\S$4, it demonstrates the effect that the shape function holds over the invariants. 

\newpage

\section{Invariants}
\subsection{Invariants for the Morris-Thorne Wormhole}
\vspace{-0.9cm}
\begin{align*}
R &= \frac{1}{r^2}\left(b'\left(r \right)\left(r \Phi'\left(r \right)+2\right)+2r \left(b\left(r \right)-r \right)\Phi''\left(r \right)-2r \left(r-b\left(r \right)\right)\Phi'\left(r \right)^2+\left(3b \left(r \right)-4r \right)\Phi'\left(r \right)\right), \tag{B.1} \label{eq:B.1} \\ \\
r_1 &= \frac{1}{16r^6}\bigl(r^2 \bigl(b' \left(r \right)^2 \left(r^2 \Phi'\left(r \right)^2+2\right)-4r b'\left(r \right)\Phi'\left(r \right) \left(r^2 \Phi''\left(r \right)+r^2 \Phi'\left(r \right)^2-2 \right) \\
& \ \ \ \ \ \ \ \ \ \ \ +4r^2\bigl(r^2 \Phi''\left(r \right)^2+r^2\Phi'\left(r \right)^4+2\Phi'\left(r \right)^2\left(r^2\Phi''\left(r \right)+1\right)\bigr)\bigr) \\ 
& \ \ \ \ \ \ \ \ \ \ \ -2 r b\left(r\right) \bigl(b'\left(r \right)\left(-2r^3\Phi' \left(r  \right)^3+\Phi'\left(r \right)\left(6r-2r^3\Phi ''\left(r \right)\right)+r^2\Phi'\left(r \right)^2+2\right) \\ 
& \ \ \ \ \ \ \ \ \ \ \ + 2r\bigl(2r^3\Phi'\left(r\right)^4+2\Phi'\left(r\right)^2 \left(2r^3\Phi''\left(r\right)+r\right)+2r\Phi''\left(r\right) \left(r^2\Phi''\left(r \right)-1\right) \\
& \ \ \ \ \ \ \ \ \ \ \ -r^2\Phi'\left(r \right)^3+\Phi'\left(r \right)\left(2-r^2 \Phi''\left(r \right)\right)\bigr)\bigr)+b\left(r\right)^2\bigl(4 r^4\Phi''\left(r \right)^2+4r^4\Phi'\left(r \right)^4 \\ 
& \ \ \ \ \ \ \ \ \ \ \ -4r^3\Phi '\left(r \right)^3-8r^2\Phi''\left(r \right)-4r\Phi'\left(r \right)\left(r^2 \Phi''\left(r \right)-3\right)+\Phi'\left(r \right)^2\left(8r^4\Phi ''\left(r \right)+r^2\right)+6\bigr)\bigr), \tag{B.2} \label{eq:B.2}\\ \\
r_2 &= -\frac{3}{64r^9}\bigl(b\left(r \right)\left(2r\Phi'\left(r \right)+1\right)-r\left(b\left(r \right)+2r\Phi'\left(r \right)\right)\bigr)^2 \\
& \ \ \ \ \ \ \ \ \ \ \ \ \bigl(r^2\left(b'\left(r\right)\Phi'\left(r\right)-2r\left(\Phi''\left(r\right)+\Phi'\left(r\right)^2\right)\right)+b\left(r\right)\left(2r^2\Phi''\left(r\right)+2r^2\Phi'\left(r\right)^2-r\Phi'\left(r\right)-2\right)\bigr), \tag{B.3} \label{eq:B.3} \\ \\
w_2 &= \frac{1}{144 r^9}\bigl(r \left(b'\left(r \right) \left(1-r \Phi'\left(r \right)\right)+2r\left(r\Phi''\left(r \right)+r\Phi '\left(r \right)^2-\Phi'\left(r \right)\right)\right) \\ 
& \ \ \ -b\left(r \right) \left(2 r^2 \Phi''\left(r \right)+2r^2 \Phi'\left(r\right)^2-3r\Phi'\left(r \right)+3\right)\bigr)^3. \tag{B.4} \label{eq:B.4}
\end{align*}
\newpage
\subsection{Invariants for the Thin Shell Schwarzschild Wormhole}
\vspace{-0.9cm}
\begin{align*}
R& = r_1 = r_2 = 0, \tag{B.5} \label{eq:B.5}\\ \\
w_2& = -\frac{12 M^3}{r^9}+\frac{6M^2}{a^2r^9}\sqrt{1-\frac{2M}{a}}\left(a \left(r-2 M\right)+2M\left(2M+2r^3-r\right)\right)\delta\left(r-a\right) \\
& \ \ \ +\frac{12M}{a^5r^9}\left(a-2M\right)\left(4M^2\left(a-2M\right)^2+r^2\left(a-2M\right)^2-4Mr\left(a-2M\right)^2-2M^2r^6\right) \delta\left(r-a\right)^2 \\
& \ \ \ \frac{8}{a^6r^9}\left(1-\frac{2 M}{a}\right)^{3/2} \left((a-2 M)^3 \left(r-2M\right)^3+M^3r^9\right)\delta\left(r-a\right)^3, \tag{B.6} \label{eq:B.6} \\ \\
w_2&|_{r=a} = \frac{2}{a^{14}} \bigl( -6a^5M^3+4a^8 \left(\frac{\left(a-2M\right)^6}{a^9}+M^3\right)\left(1-\frac{2 M}{a}\right)^{\frac{3}{2}} \\
&\ \ \ \ \ \ \ \ \ \ \ \ \ \ \ \ +3a^3M^2\left(4a^3M+a^2-4aM+4 M^2\right)\sqrt{1-\frac{2 M}{a}} \\
&\ \ \ \ \ \ \ \ \ \ \ \ \ \ \ \ -6M\left(a-2M\right)\left(2a^6 M^2-a^4+8a^3M-24a^2M^2+32aM^3-16M^4\right) \bigr). \tag{B.7}
\label{eq:B.7} \\ \\
\end{align*}
\subsection{Invariants for the Exponential Metric}
\vspace{-0.9cm}
\begin{align*}
R &= -\frac{2 M^2 e^{-\frac{2 M}{r}}}{r^4}, \tag{B.8} \label{eq:B.8} \\
r_1 &= \frac{3 M^4 e^{-\frac{4 M}{r}}}{4 r^8}, \tag{B.9} \label{eq:B.9} \\
r_2 &= \frac{3 M^6 e^{-\frac{6 M}{r}}}{8 r^{12}}, \tag{B.10} \label{eq:B.10} \\
w_2 &= -\frac{32 M^3 e^{-\frac{6 M}{r}} (2 M-3 r)^3}{9 r^{12}}. \tag{B.11} \label{eq:B.11} \\ \\
\end{align*}


\newpage

\begin{thebibliography}{1}
		
	\bibitem{1} Henry, R.~C., Overduin, J.~and Wilcomb K.~$(2016)$, ``A New Way to See Inside Black Holes,'' arXiv:1512.02762v2 [gr-qc].
	
	\bibitem{Morris:1988cz} 
  M.~S.~Morris and K.~S.~Thorne,~$(1988)$
  ``Wormholes in spacetime and their use for interstellar travel: A tool for teaching general relativity,''
  \emph{Am. J. Phys.}  {\bf 56}, 395.
  doi:10.1119/1.15620

	\bibitem{3} Morris, M.~S., Thorne, K.~S.~and Yurtsever, U.~$(1988)$, ``Wormholes, time machines, and the weak energy conditions,'' \emph{Phys.~Rev.~Lett.}, Vol.~$61$, pp.~$1446-1449$.
	
	\bibitem{4} Ellis, H.~G.~$(1973)$, ``Ether flow through a drainhole: A particle model in general relativity,'' \emph{J.~Math.~Phys.}, Vol.~$14$, pp.~$104-118$.
	
	\bibitem{5} Bronnikov, K.~A.~$(1973)$, ``Scalar-tensor theory and scalar charge,'' \emph{Acta Physica Polonica}, Vol.~B$4$, pp.~$251-266$.
	
	\bibitem{6} Visser, M.~$(1995)$, \emph{Lorentzian Wormholes: From Einstein to Hawking} (New York: AIP Press).
	
	\bibitem{7} Christoffel, E.~B.~$(1869)$, ``Ueber die Transformation der homogenen Differentialausdrücke zweiten Grades,'' \emph{Journal für die reine und angewandte Mathematik}, Vol.~$70$, pp.~$46-70$.
	
	\bibitem{8} Zakhary, E.~and McIntosh, C.~B.~G.~$(1997)$, ``A Complete Set of Riemann Invariants,'' \emph{Gen.~Relativ.~Gravit.}, Vol.~$29$, pp.~$539-581$.
	
	\bibitem{CM} Carminati, J.~and McLenaghan, R.~G.~$1991$ ``Algebraic invariants of the Riemann tensor in a four‐dimensional Lorentzian space'' \emph{Journal of Mathematical Physics}, Vol.~$32$, Num.~$11$, pp.~$3135-3140$,
    doi:10.1063/1.529470.
	
	\bibitem{Santosuosso:1998he} 
    K.~Santosuosso, D.~Pollney, N.~Pelavas, P.~Musgrave and K.~Lake,
    ``Invariants of the Riemann tensor for class B warped product spacetimes,''
    \emph{Comput. Phys. Commun.}  {\bf 115}, 381 (1998)
    doi:10.1016/S0010-4655(98)00134-9
    [gr-qc/9809012].
    
    \bibitem{Boonserm:2018orb} 
     P.~Boonserm, T.~Ngampitipan, A.~Simpson and M.~Visser,
     ``Exponential metric represents a traversable wormhole,''
     \emph{Phys. Rev. D} {\bf 98}, no. 8, 084048 (2018)
     doi:10.1103/PhysRevD.98.084048
     [arXiv:1805.03781 [gr-qc]].
	
	\bibitem{Lobo:2017oab} F.~S.~N.~Lobo~$(2017)$, ``Wormhole Basics,'' 
   in \textit{Wormholes, Warp Drives and Energy Conditions},
  Fund.\ Theor.\ Phys.\  {\bf 189}, ed. F.~S.~N.~Lobo, Springer, Cham, CH, pp. 11-33.
  doi:10.1007/978-3-319-55182-1
  
    \bibitem{McNutt:2017paq} 
     D.~Brooks, M.~A.~H.~MacCallum, D.~Gregoris, A.~Forget, A.~A.~Coley, P.~C.~Chavy-Waddy and D.~D.~McNutt,~$(2018)$
     ``Cartan Invariants and Event Horizon Detection, Extended Version,''
     \emph{Gen. Relativ. Grav.}  {\bf 50}, no. 4, 37
     [arXiv:1709.03362 [gr-qc]].
  
    \bibitem{12} Alcubierre, M. (1994), “The warp drive: hyper-fast travel within general relativity,” \emph{Class. Quant. Grav.}, Vol. 11, No. 5, pp. L73-L77.
	
	\bibitem{9} MacCallum, M.~A.~H.~$(2015)$, ``Spacetime invariants and their uses,'' arXiv:1504.06857v1 [gr-qc].
	
	\bibitem{11} D'Inverno, R.~C.~$(1992)$, \emph{Introducing Einstein's Relativity} (Oxford: University Press)
	
	
\end{thebibliography}
\end{document}